\documentclass[sigplan,screen,nonacm,pbalance]{acmart}

\usepackage{graphicx}
\usepackage{colortbl}
\usepackage{algorithm}
\usepackage{algpseudocode}

\setcopyright{acmcopyright}
\copyrightyear{2026}
\acmYear{2026}
\acmDOI{XXXXXXX.XXXXXXX}

\acmConference[CGO]{Make sure to enter the correct
  conference title from your rights confirmation emai}{February XXth--March XXth,
  2026}{Somewhere, Some country}
\acmISBN{978-1-4503-XXXX-X/18/06}

\begin{document}

\title{Memory Allocation for Constant-Bounded Programs}

\author{Vinicius Silva}
\affiliation{
  \institution{UFMG}
  \state{Minas Gerais}
  \city{Belo Horizonte}
  \country{Brazil}}
\email{vinicius.silva@dcc.ufmg.br}

\author{Kael Soares}
\affiliation{
  \institution{UFMG}
  \city{Belo Horizonte}
  \state{Minas Gerais}
  \country{Brazil}}
\email{kael@dcc.ufmg.br}

\author{M\'{a}rcio Costa}
\affiliation{
  \institution{UFMG}
  \city{Belo Horizonte}
  \state{Minas Gerais}
  \country{Brazil}}
\email{marciocs@dcc.ufmg.br}

\author{Fernando Magno Quintão Pereira}
\affiliation{%
  \institution{UFMG}
  \city{Belo Horizonte}
  \state{Minas Gerais}
  \country{Brazil}}
\email{fernando@dcc.ufmg.br}

\begin{abstract}
This work studies memory allocation for \emph{constant-bounded programs}, whose execution length is syntactically limited for all inputs.
Examples of such programs include verified kernel extensions, cryptographic routines, and fixed-shape machine-learning models.
We show that constant boundedness enables a tight, polynomial-time approximation of optimal stack usage by viewing control flow as a tree and applying a tree-scan allocation strategy augmented with memory defragmentation. Our approach guarantees memory usage bounded by the maximum live memory plus, at most, the size of the largest buffer, and is optimal when in-place swapping is permitted.
We deploy the proposed allocator in two scenarios.
First, in an Elixir-to-eBPF compiler, as a spiller that optimizes stack space.
Second, as a static heap allocator for bounded MLIR programs using the Structured Control-Flow dialect.
Results demonstrate stack reductions exceeding 90\% on real eBPF workloads and show that, even under aggressive code expansion, defragmentation is rarely required and memory usage remains a small fraction of that required by naive allocation strategies.
\end{abstract}



\maketitle

\section{Introduction}
\label{sec_intro}

A program is said to be \emph{constant-bounded} if there exists a constant $C$ such that, for all inputs, it executes at most $C$ computation steps.
That is, its running time is uniformly bounded and independent of the input.
This notion corresponds to a strong, syntactic form of termination and can be viewed as a restriction of primitive recursion in which all recursive unfoldings and loop iterations are globally bounded.
Classical examples include constant-bounded LOOP programs~\cite{Meyer67}, finite unrollings of primitive recursive definitions~\cite{Cook23}, and strongly normalizing programs whose reduction depth is bounded \emph{a priori}~\cite{Paolini06}.
Several families of real-world programs also fall into this category, including computational graphs implementing fixed-shape neural networks~\cite{Blondel25,Scarselli09}, cryptographic functions designed to be operation-invariant~\cite{Soares23}, and verified eBPF programs.

\paragraph{Contribution: Tight Approximation.}
Boundedness has positive implications for compiler optimization.
Many combinatorial problems that are NP-complete in the general case become tractable when restricted to this setting.
Since control flow can be fully unrolled, all execution paths are finite and known, and the resulting optimization search spaces are bounded.
Building on this observation, this paper presents a polynomial-time approximation for memory allocation in constant-bounded programs.
We state this result as follows:

\begin{quote}
A constant-bounded program $P$ can be compiled with at most $K = \emph{MaxLive} + \emph{MaxVar}$ memory slots, where \emph{MaxLive} is the maximum size occupied by buffers alive at any program point of $P$, and \emph{MaxVar} is the size of the largest buffer in $P$.
If swapping memory regions without an auxiliary is allowed, then $K = \emph{MaxLive}$.
\end{quote}

This result follows from the observation that the control-flow graph of a constant-bounded program can be expanded into a tree, which is amenable to tree-scan allocation, a technique well known in register allocation~\cite{Colombet11}.
Yet, unlike \citeauthor{Colombet11}'s classical approach, general allocation must handle buffers of varying sizes, a constraint that renders the problem NP-complete even for interval graphs~\cite{Lee07,Lee08}.
In this setting, a tight approximation requires \emph{defragmentation}: the ability to reallocate buffers on the stack.
Defragmentation is available in languages that support \emph{compacting garbage collection}~\cite{Cohen81}, including Java, C\#, Python, JavaScript, and Go.

\paragraph{Results: A Practical Memory Allocator.}
Section~\ref{sec_imp} describes a memory allocator that applies to any constant-bounded program, and Section~\ref{sec_eval} evaluates it on \emph{verified eBPF programs}.
The Extended Berkeley Packet Filter (eBPF) is a low-level virtual instruction set integrated into the Linux kernel~\cite{Sharaf22}, commonly used for networking, monitoring, and security tasks~\cite{Bolaji24}.
Before execution, Linux enforces a static verification to ensure that the total number of instructions executed along any program path is finite~\cite{Mohamed23,Vishwanathan23}, making eBPF an ideal target for our allocator.
We implemented the allocator in Honey Potion~\cite{Soares25}, an Elixir-to-eBPF compiler.
Section~\ref{sub_rq1} evaluates its impact on real-world eBPF programs.
Compared to the original allocator, the proposed algorithm reduces stack usage by more than 90\%, and enables the compilation of programs that could not be previously handled.

However, eBPF programs tend to be simple and do not fully exercise the algorithm.
To evaluate it more extensively, we ported the allocator to programs written in the Multi-Level Intermediate Language (MLIR)~\cite{Lattner21,Lucke25}.
Sections~\ref{sub_rq2}-\ref{sub_rq6} evaluate the algorithm on large, randomly generated programs built on top of MLIR’s Structured Control Flow dialect, enabling an analysis of the exponential impact of full code expansion.
Even under extreme conditions, we show that (i) defragmentation remains an unlikely necessity, (ii) the allocator typically uses only a fraction of the memory space required by no allocation strategy; and (iii) expanded programs tend to run faster than their original counterparts.

\section{Overview}
\label{sec_ovf}

This section provides an overview of how the proposed memory allocator operates, illustrated through a sequence of examples.
Section~\ref{sec_imp} subsequently presents the algorithms that implement each of these program transformations and memory assignments.
The proposed approach proceeds in two phases.
The first phase consists of the pre-processing steps described in Section~\ref{sub_prep}.
Section~\ref{sub_allocation} presents the second phase, namely memory allocation proper, in which variables are mapped to stack slots.

\subsection{Preprocessing}
\label{sub_prep}

Pre-processing comprises three transformations: loop unrolling,  \emph{treeification} and dead-code elimination.
Loop unrolling and dead-code elimination are well-known program transformations.
In constant-bounded programs, unrolling can be applied exhaustively, since the bounds of every loop are syntactically known.
Example~\ref{ex_unrolling} illustrates this transformation.

\begin{figure*}[ht]
\centering
\includegraphics[width=\textwidth]{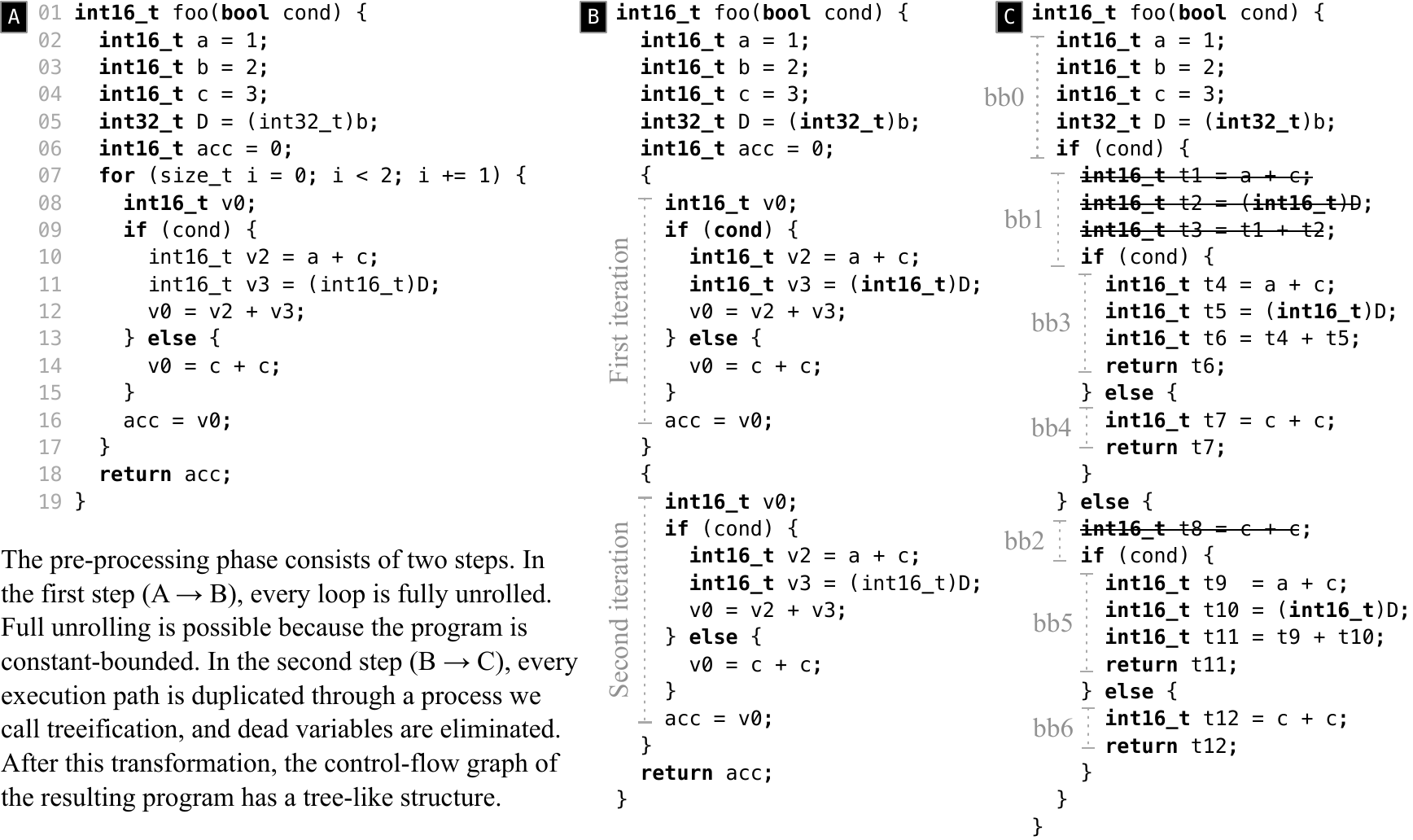}
\caption{Pre-processing phase: (1) original program; (2) program after loop unrolling; and (3) treeified program.
The final version highlights dead code (The effects of dead-code elimination on performance are discussed in Section~\ref{sub_rq5}).}
\Description{The pre-processing phase.}
\label{fig_runningCExample}
\end{figure*}

\begin{example}
\label{ex_unrolling}
Figure~\ref{fig_runningCExample}(A) shows a constant-bounded C program%
\footnote{Figure~\ref{fig_runningCExample} uses C syntax for clarity; however, the two implementations of the memory allocator evaluated in Section~\ref{sec_eval} operate on low-level program representations, namely MLIR and Elixir’s abstract syntax tree.}
that uses variables of two different sizes: two bytes (\texttt{int16\_t}) and four bytes (\texttt{int32\_t})%
\footnote{We use scalar variables allocated on the stack for illustration; the same principles apply to general memory buffers. In particular, the MLIR allocator evaluated in Section~\ref{sec_eval} maps heap-allocated buffers to stack locations.}.
The first step toward memory allocation is unrolling the loop, whose trip count is syntactically determined by Line 07 in Fig.~\ref{fig_runningCExample} (A).
Figure~\ref{fig_runningCExample} (B) shows the fully unrolled version of the program.
\end{example}

\noindent
\textbf{Syntactic Termination.}
Example~\ref{ex_unrolling} highlights an important characteristic of the class of languages considered in this paper: \emph{termination is a syntactic property}.
That is, loops and recursion are bounded by syntactically known constants.
In Example~\ref{ex_unrolling}, these constants determine the trip count of the loop, as in
``\texttt{for (i = 0; i < 2; i += 1)}''.
Section~\ref{sec_eval} evaluates two implementations of the allocator that differ only in the target language.
The first operates on Elixir programs, where iteration is implemented via recursion using a technique known as \emph{petrol semantics}~\cite{McBride15}.
In this style, recursive calls are prefixed with their maximum recursion depth, for example,
``\texttt{X = fuel 2, fact(Y)}'', where the constant \texttt{2} bounds the recursion depth.
The second implementation targets MLIR programs, in which loops use only literal bounds, such as
``\texttt{scf.for \%i = \%c0 to \%c2 step \%c1}'', where \texttt{\%c0}, \texttt{\%c1}, and \texttt{\%c2} are compile-time constants.

\paragraph{Treeification}
The second pre-processing transformation is \emph{treeification}, which fully expands all execution paths in the program.
Paths arise from conditional branches.
After expansion, the program’s control-flow graph is cycle-free and thus forms a tree, as shown in Example~\ref{ex_treeification}.

\begin{figure}[ht]
\centering
\includegraphics[width=\columnwidth]{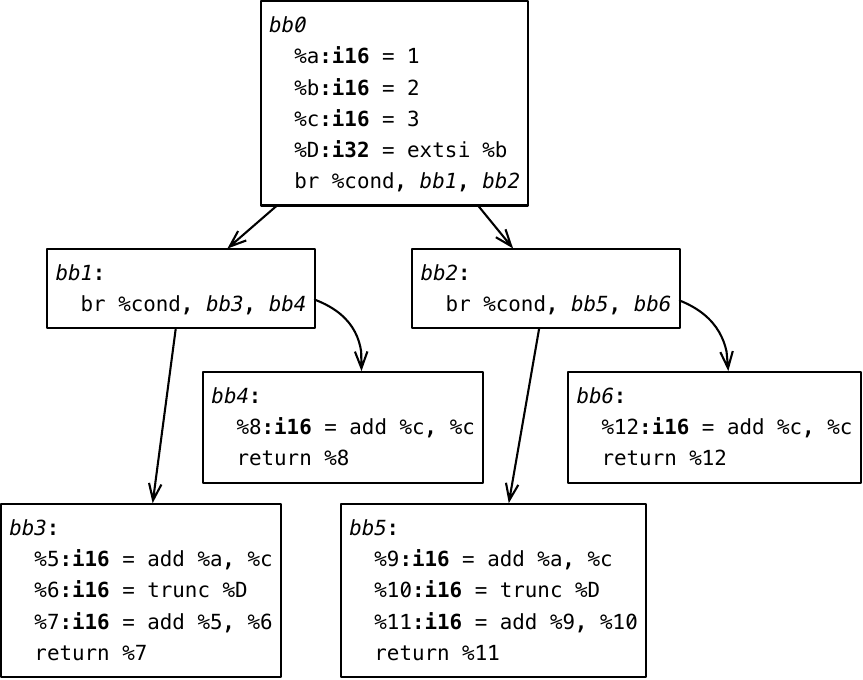}
\caption{Treeified control-flow graph of Figure~\ref{fig_runningCExample} (C).}
\Description{Treeified control-flow graph.}
\label{fig_treeifiedCFG}
\end{figure}

\begin{example}
\label{ex_treeification}
Figure~\ref{fig_runningCExample} (C) shows the treeified version of the unrolled program from Figure~\ref{fig_runningCExample} (B).
The primary effect of treeification is the duplication of execution paths, resulting in a tree-shaped control-flow graph.
Figure~\ref{fig_treeifiedCFG} depicts the control-flow graph of the treeified program.
Treeification also duplicates program variables.
New variables receive fresh names, making the transformed program trivially in static single assignment (SSA) form~\cite{Cytron91}.
In this example, each variable has a single definition that dominates all its uses.
\end{example}

\noindent
\textbf{Dead-Code Elimination.}
Although not a requirement for memory allocation, we perform dead-code elimination after treeification.
The elimination of dead variables is trivial in treeified programs, because they have the static single-assignment property, and do not contain cyclic uses of variables (the SSA-form treeified program does not contain $phi$-functions).
Thus, a variable is dead if it: (base case) has no uses, or (inductive case) it is only used in the definition of dead variables, as Example~\ref{ex_dead_elim} shows.

\begin{example}
\label{ex_dead_elim}
Figure~\ref{fig_runningCExample} (C) highlight dead variables.
Variables \texttt{t1}, \texttt{t2} and \texttt{t3} have no uses; hence, fit into the base definition of dead code.
Variable \texttt{t2} is only used in the definition of \texttt{t3}; hence, fit into the inductive definition.
\end{example}

\subsection{Tree-Scan Allocation}
\label{sub_allocation}

Memory allocation is performed on the treeified control-flow graph (Figure~\ref{fig_runningCExample} in Example~\ref{ex_treeification}).
Allocation uses a \emph{tree-scan} algorithm, a generalization of the classic linear scan allocator~\cite{Poletto99} to tree-structured control flow.
The optimality of the proposed approach relies on Theorem~\ref{theo_global_bound} (Page~\pageref{theo_global_bound}), which states that if allocation is optimal along every root-to-leaf path of the tree, then it is optimal for the entire tree.
Achieving optimality along a path requires a process called \emph{defragmentation}, illustrated in Example~\ref{ex_defragmentation}.

\begin{figure}[ht]
\centering
\includegraphics[width=\columnwidth]{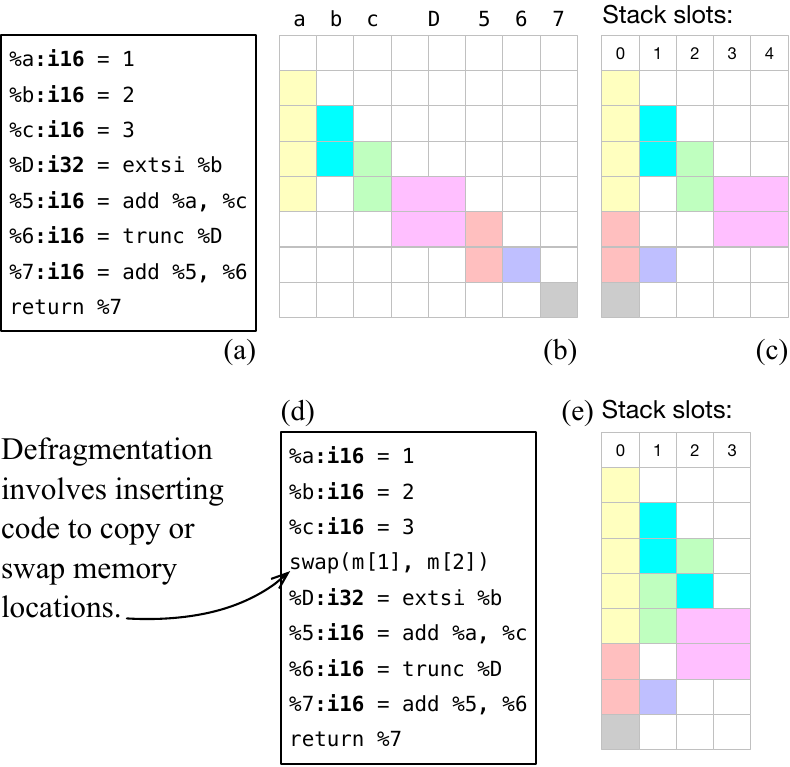}
\caption{
(a) Linearized path $\mathtt{bb0} \rightarrow \mathtt{bb1} \rightarrow \mathtt{bb3}$ from Figure~\ref{fig_treeifiedCFG}.
(b) Instance of the memory allocation problem.
(c) Greedy allocation.
(d) Code after insertion of defragmentation routines.
(e) Allocation after defragmentation.}
\Description{Example of allocation.}
\label{fig_allocation1}
\end{figure}

\begin{example}
\label{ex_defragmentation}
Figure~\ref{fig_allocation1} (a) shows a linearized path extracted from the tree in Figure~\ref{fig_treeifiedCFG}.
In this context, memory allocation is equivalent to packing the live intervals shown in Figure~\ref{fig_allocation1} (b) as tightly as possible.
In this example, most variables occupy two bytes on the stack, while one variable occupies four bytes.
Greedy allocation is suboptimal: Figure~\ref{fig_allocation1} (c) shows that it leaves holes in the stack layout, notably around variable \texttt{D}.
Defragmentation eliminates these holes by inserting additional instructions that move variables in memory.
One such instruction, a memory swap, is shown in Figure~\ref{fig_allocation1} (d).
After defragmentation, the stack contains no empty slots; variables occupy a contiguous block of addresses, as illustrated in Figure~\ref{fig_allocation1} (e).
\end{example}

Lemma~\ref{lem_path_optimality} (Page~\pageref{lem_path_optimality}) proves that path-wise allocation with defragmentation is optimal.
Intuitively, once defragmentation is allowed, the maximum required stack size equals the maximum total size of simultaneously live variables.
However defragmentation is not desirable: it inserts instructions on the target program to implement memory shuffling and prevents loop reroling for code-size compression~\cite{Rocha22}.
Section~\ref{sec_eval} shows, empirically, that defragmentation is rare.
Minimizing it is not a goal of this paper; thus, we leave open the problem of reducing the number defragmentation points.

\section{Implementation}
\label{sec_imp}

\begin{figure}[ht]
\centering
\includegraphics[width=\columnwidth]{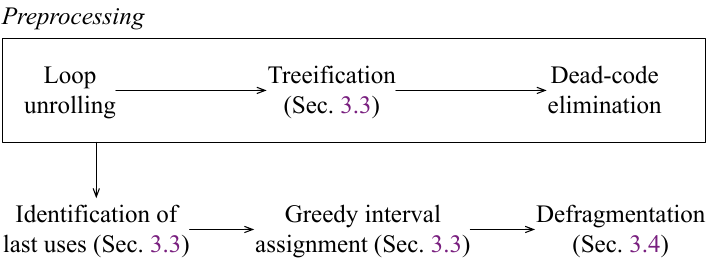}
\caption{Overview of the proposed memory allocator.}
\Description{Overview of the memory allocation.}
\label{fig_algorithmic_overview}
\end{figure}

Figure~\ref{fig_algorithmic_overview} presents an overview of the new memory allocator.
Memory allocation, as proposed in this paper, is \emph{fully static}: the compiler knows the address of every memory buffer.
Nevertheless, this approach can be applied to settings such as stack and heap allocation, which are traditionally considered dynamic.
To demonstrate this point, Section~\ref{sec_eval} evaluates a spiller (i.e., a stack allocator) for eBPF programs, as well as a heap allocator for MLIR programs.
Static allocation is enabled by restricting the allocator to constant-bounded programs, which are amenable to the treeification process outlined in Section~\ref{sub_prep} and described in full detail in Section~\ref{sub_treeify}.

\subsection{Preprocessing - Treeification}
\label{sub_treeify}

As already explained in Section~\ref{sub_prep}, preprocessing consists of three transformations: loop unrolling, treeification and dead-code elimination.
Only treeification is non-standard; hence, we omit discussing the other two pre-processing steps.
Treeification, a process that Algorithm~\ref{alg_treeification} represents, transforms a control-flow graph (CFG) into a tree by duplicating shared control-flow paths.
Starting from the entry instruction, Algorithm~\ref{alg_treeification} recursively clones each instruction and its successors.
Branch instructions recursively duplicate both successors, while sequential instructions duplicate their single successor.
As a result, all join points are eliminated, yielding a tree-shaped CFG, as seen in Example~\ref{ex_treeification}.

\begin{algorithm}[htb]
\caption{CFG Treeification}
\label{alg_treeification}
\begin{algorithmic}[1]
\Function{Treeify}{$CFG = (V \cup \{\mathit{init}\}, E)$}
  \State \Return \Call{Duplicate}{$\mathit{init}$}
\EndFunction

\Function{Duplicate}{$v$}
  \State $v' \gets \Call{Clone}{v}$
  \ForAll{$u \in \Call{Successors}{v}$}
    \State add edge $(v', \Call{Duplicate}{u})$
  \EndFor
  \State \Return $v'$
\EndFunction
\end{algorithmic}
\end{algorithm}

Two remarks concerning Algorithm~\ref{alg_treeification} are in order.
First, it duplicates shared control-flow paths and therefore increases code size exponentially on the number of branches. Second, Algorithm~\ref{alg_treeification} must be applied only after loop unrolling, otherwise the recursive call in Line 7 would not terminate.

\subsection{Identification of Last Uses}
\label{sub_last_uses}

The memory allocation procedure that Section~\ref{sub_greedy_assignment} requires knowledge of the {\it last uses} of each variable.
The set $\mathit{LastUse}(v)$ of a variable $v$ within a treeified CFG $G$ is the collection of instructions $i \in G$ with the following two properties:
(i) The instruction $i$ contains a use of $v$; and (ii) there is a path from $i$ to a terminator---an instruction without successor---that does not cross any other use of $v$.
Algorithm~\ref{alg_collect_last_uses} computes last-use information for SSA values by traversing the treeified control-flow graph in a bottom-up manner.

\begin{algorithm}[htb]
\caption{Bottom-Up Identification of Last Uses on a Treeified CFG}
\label{alg_collect_last_uses}
\begin{algorithmic}[1]
\Function{CollectLastUses}{$B$}
  \State $Seen \gets \emptyset$
  \ForAll{successor blocks $S$ of $B$}
    \State $Seen \gets Seen \cup \Call{CollectLastUses}{S}$
  \EndFor

  \ForAll{operations $op$ in $B$, in reverse order}
    \ForAll{operands $v$ of $op$}
      \If{$v \notin Seen$}
        \State $LastUse(v) \gets LastUse(v) \cup \{op\}$
        \State $Seen \gets Seen \cup \{v\}$
      \EndIf
    \EndFor
  \EndFor

  \State \Return $Seen$
\EndFunction
\end{algorithmic}
\end{algorithm}

For each basic block $B$, the algorithm first recursively processes all successor blocks, collecting the set of values that are live beyond $B$.
It then scans the operations of $B$ in reverse program order.
Whenever an operand $v$ is encountered that has not been seen in any successor or later operation, the current operation is recorded as a last use of $v$.
The algorithm returns the set of values that are live at the entry of $B$.

\begin{example}
\label{ex_last_use} The set of last use sites of variable \texttt{\%b} in Figure~\ref{fig_treeifiedCFG} is $\mathit{LastUse}(\mathtt{\%b}) = \{\mathtt{\%D = extsi \ \%b} \}$.
Similarly, we have that $\mathit{LastUse}(\mathtt{\%D}) = \{\mathtt{\%6 = trunc \ \%D}, \mathtt{\%10 = trunc \ \%D} \}$.
\end{example}

\subsection{Greedy Interval Assignment}
\label{sub_greedy_assignment}

Once last-usage data has been collected, Algorithm~\ref{alg_tree_scan_alloc} performs stack allocation on a treeified control-flow graph using a top-down traversal.
Operations are processed in program order, allocating stack space for each value at its definition and freeing that space at its last use.
Freed intervals are placed in a free list and reused whenever possible.

\begin{algorithm}[htb]
\caption{Tree-Scan Stack Allocation}
\label{alg_tree_scan_alloc}
\begin{algorithmic}[1]
\Function{AllocateBlock}{$B, State$}
  \ForAll{operations $op$ in $B$, in program order}
    
    \Comment{Free operands at their last use}
    \ForAll{operands $v$ of $op$}
      \If{$op$ is a last use of $v$}
        \State \Call{Free}{$State, Active[v]$}
        \State remove $v$ from $Active$
      \EndIf
    \EndFor

    \Comment{Allocate memory space for results}
    \ForAll{results $r$ of $op$}
      \State $size \gets \Call{SizeOf}{r}$
      \State $interval \gets \Call{AllocateInterval}{State, size}$
      \State $Active[r] \gets interval$

      \Comment{Immediately free dead values}
      \If{$r$ has no uses}
        \State \Call{Free}{$State, interval$}
        \State remove $r$ from $Active$
      \EndIf
    \EndFor

  \EndFor
\EndFunction
\end{algorithmic}
\end{algorithm}

Memory assignment is handled by Algorithm~\ref{alg_allocate_interval}.
The allocator first attempts to reuse a contiguous free interval.
If this is not possible and allocating from the top of the memory would increase the maximum observed memory size, the allocator may invoke a defragmentation routine, which Section~\ref{sub_defrag} shall explain.
Otherwise, allocation proceeds by extending the amount of used memory.
The maximum memory size is tracked globally across all execution paths.

\begin{algorithm}[htb]
\caption{Interval Allocation}
\label{alg_allocate_interval}
\begin{algorithmic}[1]
\Function{AllocateInterval}{$State, size$}
  \If{\Call{HasContiguousFree}{$State, size$}}
    \State \Return \Call{AllocateFromFreeList}{$State, size$}
  \EndIf

  \If{\Call{TotalFreeMemory}{$State$} $> 0$ \textbf{and} \\
      $\mathit{State.nextOffset} + size > \mathit{GlobalMaxMemSize}$}
    \State \Call{Defragment}{$State$}
  \EndIf

  \State \Return \Call{AllocateFromTop}{$State, size$}
\EndFunction
\end{algorithmic}
\end{algorithm}

\subsection{Defragmentation}
\label{sub_defrag}

Lines 5-7 of Algorithm~\ref{alg_allocate_interval} indicate that when a new allocation would increase the maximum stack size, the allocator invokes a defragmentation routine.
The routine, represented as Algorithm~\ref{alg_defragment}, compacts all live stack intervals while preserving correctness with respect to variable lifetimes.
Active values are partitioned into two sets:
\begin{description}
\item[movable values:] Values whose lifetimes extend beyond the current program point.
\item[protected values:] Values whose last use occurs at the current point.
\end{description}
Movable values are packed first, starting from offset zero.
Protected values are then placed contiguously above them.
This {\it modus operandi} guarantees that values whose lifetimes end immediately remain at the end of the occupied memory, allowing their space to be reused by the forthcoming allocation.
Example~\ref{ex_shuffling} explains this need.

\begin{example}
\label{ex_shuffling}
In Figure~\ref{fig_allocation1} (e), defragmentation swaps the slots occupied by variables \texttt{\%b} and  \texttt{\%c}.
This shuffling ensures that \texttt{\%b}, which has an imminent last use, remains on top of the interval of allocated memory slots.
In this way, the next instruction, which is \texttt{\%b}'s last use, can reuse that variable's memory slot to allocate variable \texttt{\%D}.
\end{example}

\begin{algorithm}[t]
\caption{Defragmentation with Last-Use Awareness}
\label{alg_defragment}
\begin{algorithmic}[1]
\Require Allocator state $S$, set $L$ of values whose last use occurs at this point
\Ensure Active values packed at minimal offsets, preserving last-use values at the top

\State $M \gets \emptyset$ \Comment{Movable values}
\State $P \gets \emptyset$ \Comment{Protected (last-use) values}

\ForAll{$(v, i) \in S.\textit{active}$}
  \If{$v \in L$}
    \State $P \gets P \cup \{(v, i)\}$
  \Else
    \State $M \gets M \cup \{(v, i)\}$
  \EndIf
\EndFor

\State Sort $M$ by increasing offset
\State Sort $P$ by increasing offset

\State $o \gets 0$ \Comment{Next free offset}

\ForAll{$(v, i) \in M$}
  \If{$i.\textit{offset} \neq o$}
    \State $\textsc{Move}(v, o)$
  \EndIf
  \State $i.\textit{offset} \gets o$
  \State $o \gets o + i.\textit{size}$
\EndFor

\ForAll{$(v, i) \in P$}
  \If{$i.\textit{offset} \neq o$}
    \State $\textsc{Move}(v, o)$
  \EndIf
  \State $i.\textit{offset} \gets o$
  \State $o \gets o + i.\textit{size}$
\EndFor

\State $S.\textit{nextOffset} \gets o$
\State $S.\textit{freeList} \gets \emptyset$
\end{algorithmic}
\end{algorithm}

To implement defragmentation, Algorithm~\ref{alg_defragment} performs a sequence of logical memory moves, updates interval offsets accordingly, and resets the free list.
Thus, Algorithm~\ref{alg_defragment}'s asymptotic complexity is linear on the number of active variables at the point it is invoked.
After defragmentation, the stack prefix is fully compact, and future allocations occur at the smallest possible offset.

\section{Correctness}
\label{sec_correctness}

This section establishes the correctness and optimality guarantees of the proposed memory allocator.
Section~\ref{sub_semantics} shows that the program transformations performed by the allocator preserve semantics.
Section~\ref{sub_optimality} shows that the resulting memory layout is asymptotically optimal for constant-bounded programs.

\subsection{Semantic Preservation}
\label{sub_semantics}

Semantic preservation is presented incrementally.
We first show that the preprocessing phase preserves program semantics.
We then prove optimality of memory allocation along individual execution paths.
Finally, we lift this result to tree-shaped control-flow graphs and derive a global bound on stack usage.
The preprocessing phase of the allocator performs two transformations: loop unrolling and treeification.
Both transformations rely on the fact that the input program is constant-bounded.
The following theorem states that these transformations preserve program semantics.

\begin{theorem}[Semantic Preservation of Treeification]
\label{theo_semantic_equivalence}
For every constant-bounded program $P$, the treeified program $P_T$ is semantically equivalent to $P$.
That is, for all inputs, $P$ and $P_T$ produce the same observable results.
\end{theorem}

\begin{quote}
\begin{small}

Since $P$ is constant-bounded, all loops and recursive calls are syntactically bounded and can be exhaustively unrolled.
Treeification duplicates control-flow paths but does not alter the order, number, or semantics of executed operations along any feasible execution.
Each execution of $P$ corresponds to exactly one root-to-leaf path in $P_T$, and vice versa.
\qed

\end{small}
\end{quote}

Theorem~\ref{theo_semantic_equivalence} allows us to reason about memory allocation on $P_T$ instead of $P$.
Variable renaming introduced during treeification ensures Static Single Assignment form.
Thus, in the remainder of this section, we assume that programs are represented as tree-shaped control-flow graphs in SSA form.

\subsection{Optimality}
\label{sub_optimality}

Once the control-flow graph has been treeified, memory allocation is performed independently along each root-to-leaf path.
Along a single path, memory allocation reduces to the problem of assigning stack slots to a sequence of variable live intervals as stated in Lemma~\ref{lem_path_optimality}.
In this lemma, $\pi$ denotes a linearized execution path, and $\emph{MaxLive}(\pi)$ denotes the maximum total size of variables simultaneously alive at any point along $\pi$.

\begin{lemma}[Path Optimality with Defragmentation]
\label{lem_path_optimality}
For any linearized execution path $\pi$, the allocator with defragmentation produces a memory layout whose size is exactly $\emph{MaxLive}(\pi)$.
\end{lemma}

\begin{quote}
\begin{small}

Allocation along a path proceeds via a greedy assignment of stack slots to variables as they become live.
When variables of different sizes are present, greedy allocation alone may introduce holes in the layout.
Defragmentation eliminates such holes by relocating live variables, ensuring that live memory occupies a contiguous prefix of the stack.
As a result, at every program point, the stack size equals the total size of live variables.
\qed

\end{small}
\end{quote}

We now show, in Lemma~\ref{lem_tree_optimality}, that optimality along individual paths suffices to guarantee optimality for the entire tree-shaped control-flow graph.
Let $T$ be a tree-shaped control-flow graph, and let $\emph{MaxLive}(T)$ denote the maximum live memory over all nodes of $T$.

\begin{lemma}[Tree Optimality]
\label{lem_tree_optimality}
If memory allocation is optimal along every root-to-leaf path of a tree-shaped control-flow graph $T$, then the allocation is optimal for $T$ as a whole.
\end{lemma}

\begin{quote}
\begin{small}

In a tree-shaped control-flow graph, every execution corresponds to exactly one root-to-leaf path.
The maximum stack usage of the program is therefore the maximum stack usage along any such path.
If allocation is optimal for each path, no execution can require more than $\emph{MaxLive}(T)$ stack space.
\qed

\end{small}
\end{quote}

By combining Lemmas~\ref{lem_tree_optimality} and~\ref{lem_path_optimality} we reach the main result of the paper, namely, that the proposed memory allocator is optimal.

\begin{theorem}[Global Stack Bound]
\label{theo_global_bound}
Every constant-bounded program $P$ can be allocated with a stack of size at most
\[
\emph{MaxLive}(P) + \varepsilon,
\]
where $\varepsilon$ is a temporary memory region used to perform defragmentation.
Moreover, $\varepsilon$ can be bounded by the size of the largest variable in $P$.
\end{theorem}

\begin{quote}
\begin{small}

By Theorem~\ref{theo_semantic_equivalence}, allocation on $P_T$ is semantically equivalent to allocation on $P$.
By Lemma~\ref{lem_path_optimality}, allocation along each path of $P_T$ is optimal up to a temporary buffer $\varepsilon$.
By Lemma~\ref{lem_tree_optimality}, this optimality lifts to the entire tree.
Thus, the required stack size is bounded by $\emph{MaxLive}(P)$ plus the temporary space needed to perform defragmentation.
\qed

\end{small}
\end{quote}

The size of $\varepsilon$ depends on the underlying instruction set.
If in-place memory swaps are supported, $\varepsilon$ can be as small as the width of a machine register.
If memory movement is implemented via bulk copy operations (e.g., \texttt{memcpy}), $\varepsilon$ must be large enough to hold the largest buffer.
In all cases, $\varepsilon \leq \emph{MaxVar}(P)$.
As a corollary, when in-place swaps are available, the allocator achieves a stack size of exactly $\emph{MaxLive}(P)$.

\section{Evaluation}
\label{sec_eval}

The goal of this section is to evaluate the following research questions:
\begin{description}
\item[RQ1:] What is the efficacy of deploying the proposed memory allocator on real-world eBPF programs?
\item[RQ2:] What are the running time and asymptotic behavior of the proposed allocator?
\item[RQ3:] What is the effect of tree-scan on the memory footprint of programs?
\item[RQ4:] How likely is defragmentation to occur in large and complex programs?
\item[RQ5:] What is the impact of memory allocation on the running time of programs?
\item[RQ6:] What is the effect of treeification on the binary size of programs?
\end{description}

\paragraph{Experimental Setup:}
Experiments run on an AMD Ryzen 7 3700X (8 cores, 16 threads) with 32\,GB of RAM, featuring Ubuntu 22.04 LTS.
The memory allocator was evaluated as part of Honey Potion v0.2.0 and on MLIR version 22.0.

\noindent
\textbf{Baselines for Comparison:}
This section compares three memory allocators.
All allocators receive the same input program and produce a mapping from variables to stack slots.
The allocators considered are:
\begin{description}
\item[Trivial:] An allocator that assigns each variable to a distinct stack slot.
\item[Tree-Scan:] The allocator described in Section~\ref{sec_imp}, without defragmentation.
\item[Defrag:] The allocator described in Section~\ref{sec_imp}, with support for defragmentation.
\end{description}

\noindent
\textbf{Honey Potion Benchmarks:}
We evaluate \textbf{RQ1} using 24 eBPF programs drawn from the Honey Potion repository\footnote{Available at \url{https://github.com/lac-dcc/honey-potion}}.
Ten of these programs are real applications; the remaining programs are synthetic benchmarks designed to exercise different features of the compiler.
Even the application programs tend to be very small, reflecting the nature of eBPF workloads, which typically implement event handlers.
Consequently, with the exception of two programs, their control-flow graphs (CFGs) are already tree-shaped and do not require code expansion.
Moreover, none of these 24 benchmarks trigger defragmentation.

\noindent
\textbf{BenchGen Benchmarks:}
To investigate \textbf{RQ2--RQ6}, larger benchmarks are required.
To obtain such benchmarks, we implemented the allocator for MLIR~\cite{Lattner21} programs and used BenchGen~\cite{Silva25} to generate a corpus of large MLIR programs.
BenchGen is a benchmark generator based on L-systems: users define grammars that determine how programs grow.
By iteratively applying these grammar rules, BenchGen produces successive generations of programs, each exponentially larger than the previous one.
The grammar allows users to specify loops, conditional blocks, scope rules as well as variable definitions and uses.

Our customized BenchGen produces programs using two MLIR dialects: the Structured Control Flow (SCF) dialect with bounded loops, and the Arithmetic dialect, which supports variables of 8, 16, 32, and 64 bits.
During treeification, these programs are lowered to the Control Flow (CF) dialect, as SCF does not support tree-shaped control-flow graphs.
Memory allocation is performed on the stack and is fully static: since programs are bounded and fully unrolled, the compiler can determine the address of each variable at compile time.

\subsection{RQ1: Evaluation on eBPF Programs}
\label{sub_rq1}

\begin{figure}[ht]
\centering
\includegraphics[width=\columnwidth]{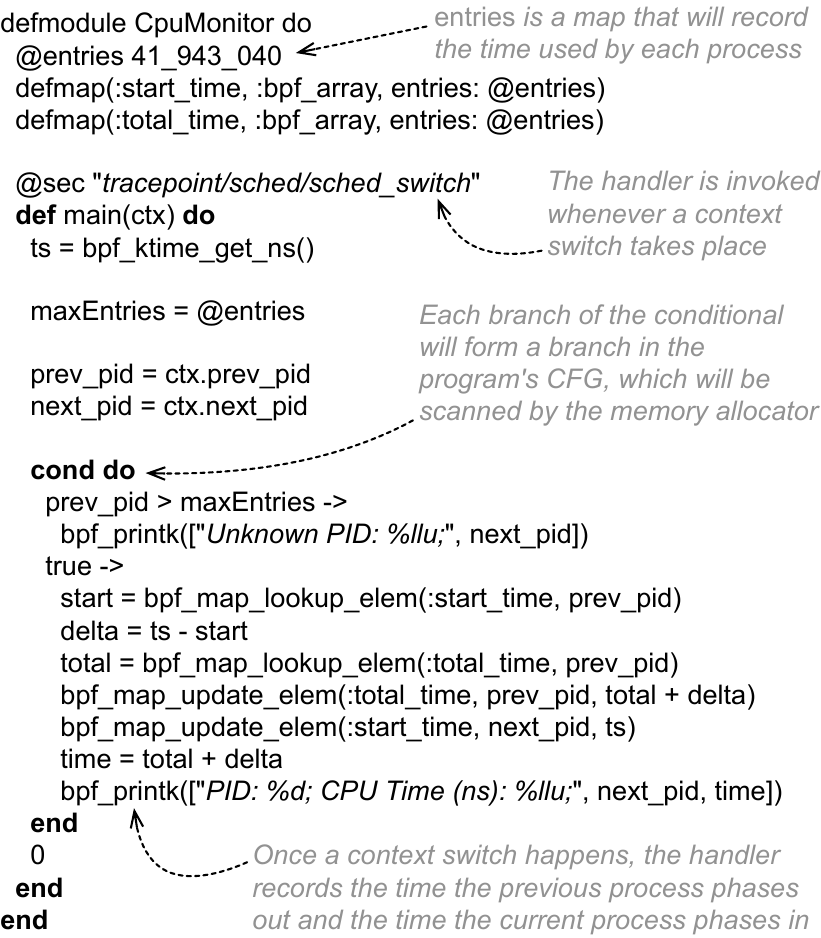}
\caption{An example of an eBPF program written in Elixir: this program monitors the time taken by each CPU process.}
\Description{An example of an eBPF program written in Elixir.}
\label{fig_cpuMonitorHoneyPotion}
\end{figure}

This section evaluates the proposed allocator within an actual eBPF compiler: Honey Potion.
As described in Section~\ref{sec_intro}, Honey Potion translates Elixir programs into verified eBPF programs.
Figure~\ref{fig_cpuMonitorHoneyPotion} shows an example of such a program.
Its control-flow graph is already tree-shaped; therefore, allocation does not require code duplication.
Indeed, this is the case for all but two of the 24 available benchmarks, which are generally short programs.
As a consequence of this simplicity, the experiments reported in this section omit the \textbf{Defrag} allocator, since defragmentation was not required for any benchmark in the Honey Potion repository.

\begin{figure}[ht]
\centering
\includegraphics[width=\columnwidth]{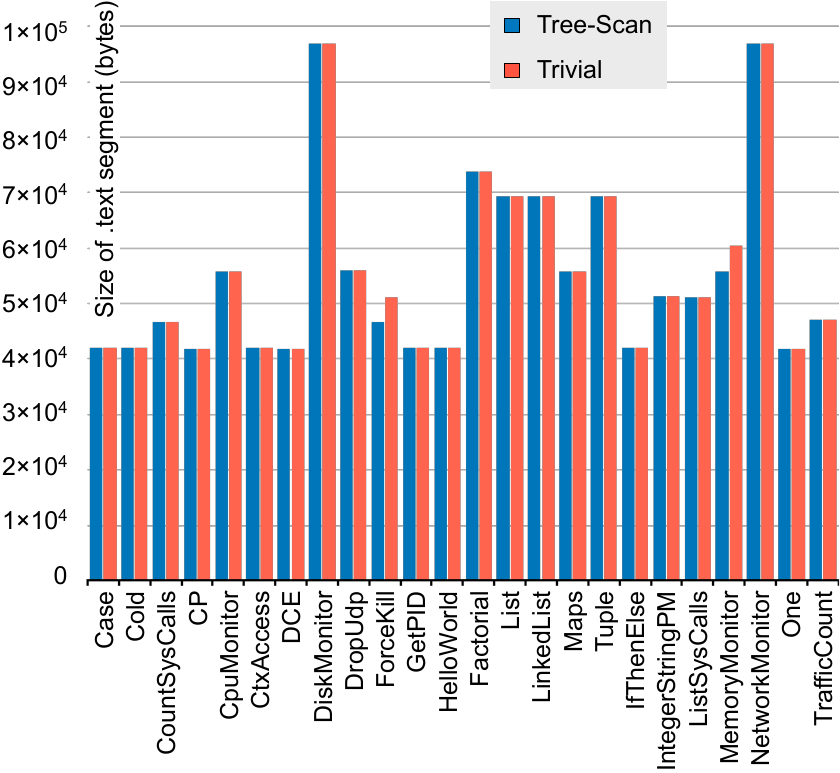}
\caption{Binary size (bytes of the \texttt{.text} segment) of eBPF programs. Differences observed for \texttt{ForceKill} and \texttt{MemoryMonitor} are caused by code duplication.}
\Description{Binary size (bytes of the \texttt{.text} segment) of the eBPF programs produced for each benchmark.}
\label{fig_binarySizeHoneyPotion}
\end{figure}

\paragraph{Discussion:}
Figure~\ref{fig_binarySizeHoneyPotion} reports the size of the binaries produced by Honey Potion using the two allocators under consideration.
Code duplication occurred in two benchmarks due to conditional expressions.
For instance, the control flow of an expression such as
``\texttt{r = if a > b, do: a, else: b}''
contains a join point at which variable \texttt{r} is defined.
Although conditional expressions are common in Elixir programs, they are rare in eBPF workloads.
These programs tend to follow the pattern illustrated in Figure~\ref{fig_cpuMonitorHoneyPotion}, where conditional branches are independent.
As a result, code duplication is both infrequent and has limited impact on binary size, accounting for less than 10\% growth in each of the two cases.

\begin{figure}[ht]
\centering
\includegraphics[width=\columnwidth]{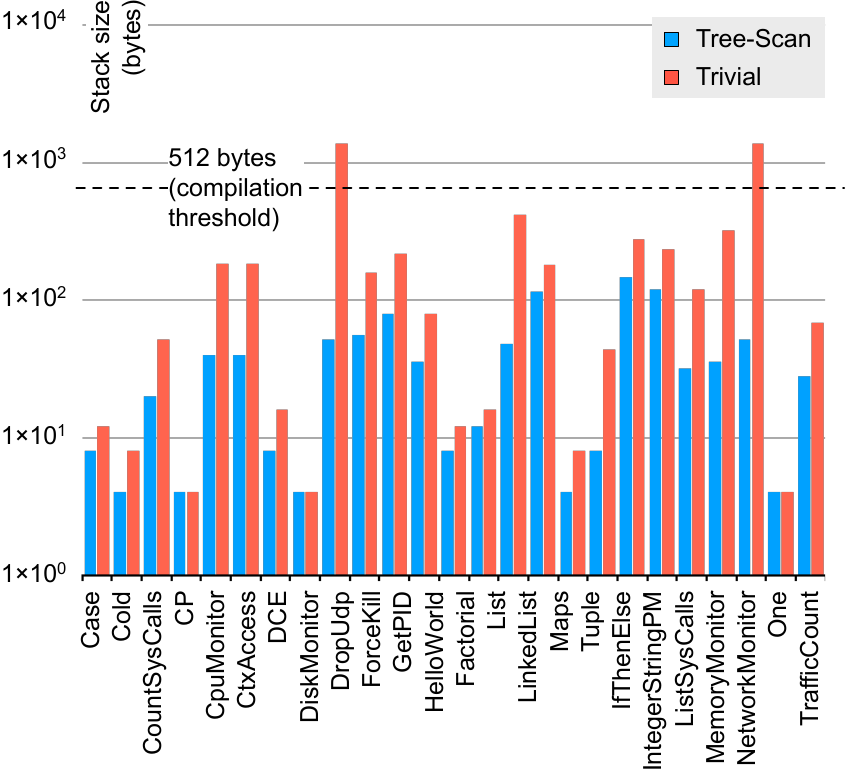}
\caption{Stack space required to compile each program. Neither \texttt{NetworkMonitor} nor \texttt{TrafficCount} can be compiled with the \textbf{Trivial} allocator, as they exceed the 512-byte limit imposed by the Linux eBPF verifier.}
\Description{Total amount of stack space required to compile each program.}
\label{fig_absoluteStackSizeHoneyPotion}
\end{figure}

The \textbf{Trivial} allocator, which assigns each variable to a separate stack slot, requires significantly more stack space, both in absolute terms and relative terms, as Figure~\ref{fig_absoluteStackSizeHoneyPotion} shows.
Two benchmarks could not be compiled with the \textbf{Trivial} allocator, as they exceed the 512-byte stack limit enforced by the Linux eBPF verifier.

\subsection{RQ2: Running Time and Asymptotic Behavior}
\label{sub_rq2}

To compare the allocation time taken by the different memory allocators, we have used six different BenchGen grammars, to produce 11 generations of MLIR programs.
This approach gives us $6 \times 11 = 66$ programs, which we also use in Sections~\ref{sub_rq3} and~\ref{sub_rq4}.
Programs from generation $n$ are about twice as large as programs from generation $n-1$, considering number of MLIR instructions and number of variables.

\begin{figure*}[ht]
\centering
\includegraphics[width=\textwidth]{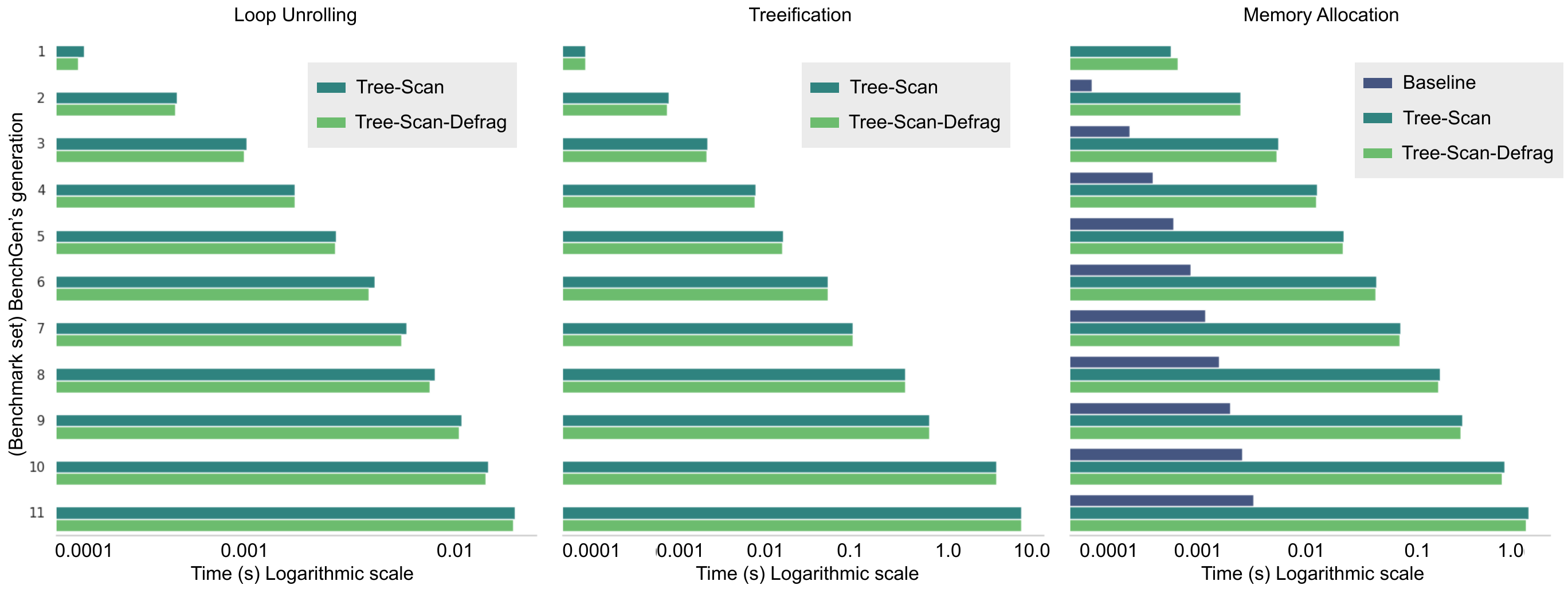}
\caption{Memory allocation time that different allocators take on the MLIR programs that BenchGen generates automatically.
Each bar shows the average time observed on a population of six programs.}
\Description{Memory allocation time that the different allocators take on the MLIR programs that BenchGen generates automatically.}
\label{fig_runtimeBenchGen}
\end{figure*}

\paragraph{Discussion:}
Figure~\ref{fig_runtimeBenchGen} subsumes our observations.
The most expensive transformation is treeification.
It takes about 10 seconds to treeify all the six programs in the 11-th generation of benchmarks.
These programs contains about $10^4$ variables.
The memory allocation of treeified programs runs in about one tenth of the time necessary to run treeification.
Loop unrolling, in turn, is much faster, running on about one-hundredth of the time it takes to do memory allocation.
In other words, loop unrolling seems to be two orders of magnitude faster than memory allocation (on treeified programs), which seems to be one order of magnitude faster than treeification.

This experiment indicates that memory allocation of treeified programs is about 20x slower than memory allocation of the original programs.
This factor is less than he size ratio between original and treeified programs.
On the 11-th generation, treeified programs are about 60x larger than the original programs.
Memory allocation, with or without defragmentation, is linearly proportional to the size of the treeified program: the coefficient of determination ($R^2$) is 0.99, almost a perfect correlation.

\subsection{RQ3: Effect on Memory Footprint}
\label{sub_rq3}

In the MLIR setting, BenchGen generates scalar variables with different bitwidths.
These variables are mapped to stack slots by the memory allocators.
Thus, the more effective the allocator, the smaller the maximum stack usage it requires.

\begin{figure}[ht]
\centering
\includegraphics[width=\columnwidth]{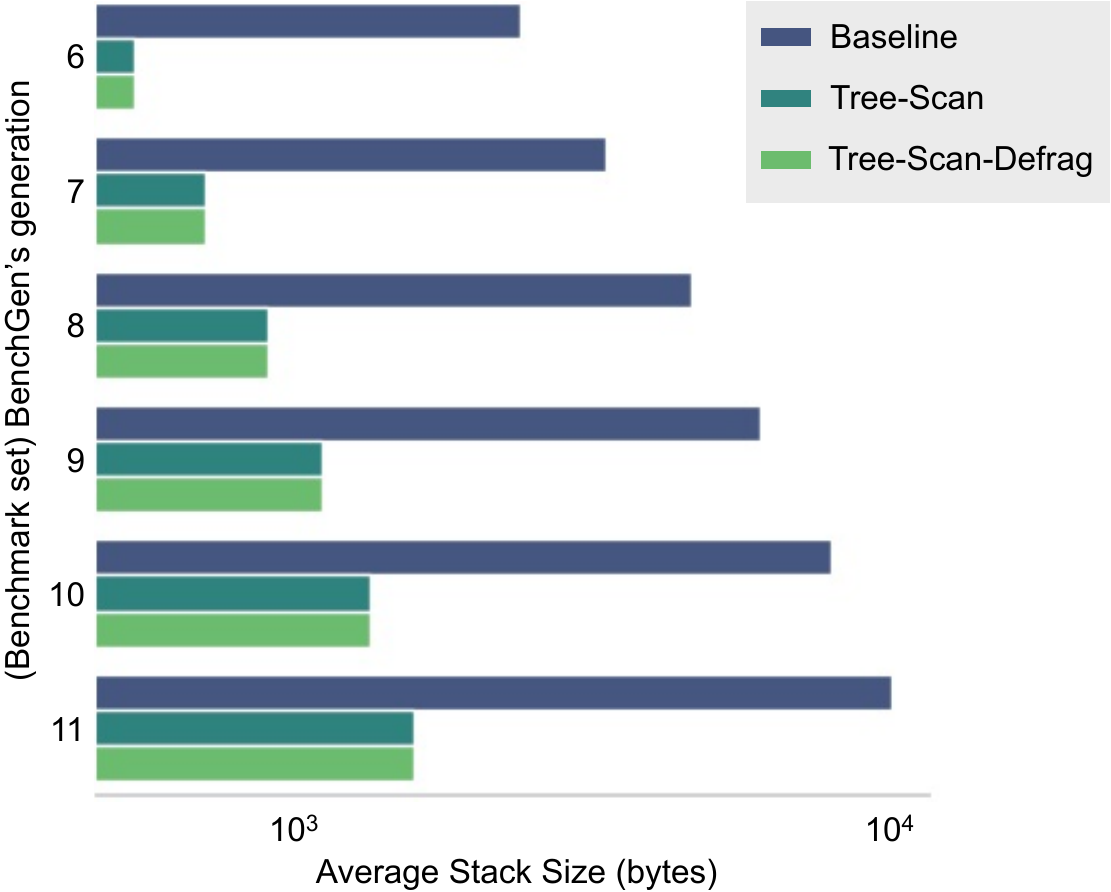}
\caption{Average stack space required to compile programs.
Each bar represents the average of six samples.}
\Description{Average stack space required to compile programs.}
\label{fig_averageStackSize}
\end{figure}

\paragraph{Discussion:}
Figure~\ref{fig_averageStackSize} shows the average stack space required by each allocator.
Tree-scan allocation is one order of magnitude more efficient than the baseline approach.
For the largest benchmarks (six 11th-generation programs), plain tree-scan uses 11,892 bytes in total.
With defragmentation, allocation requires 8,774 bytes of stack.
In contrast, the baseline allocator requires 82,138 bytes, representing a growth of 6.9$\times$.
In spite of its impact, defragmentation is rarely needed in this benchmark suite, as Section~\ref{sub_rq4} shows.

\subsection{RQ4: Analysis of Defragmentation}
\label{sub_rq4}

As already observed in Section~\ref{sub_rq1} and further suggested in Section~\ref{sub_rq3}, defragmentation is a rare phenomenon—at least for the benchmarks evaluated in this paper.
This section quantifies how rare it is.
To this end, we report three metrics related to defragmentation:
\begin{description}
\item[\#Defrag:] The number of times the defragmentation routine (Algorithm~\ref{alg_defragment}) is invoked.
\item[\#Moves:] The total number of logical moves performed by Algorithm~\ref{alg_defragment} (Lines 15 and 22).
\item[\#Bytes:] Number of bytes moved due to defragmentation (Sum of offsets evaluated to true in Lines 14 and 21).
\end{description}

\begin{figure}[ht]
\centering
\includegraphics[width=\columnwidth]{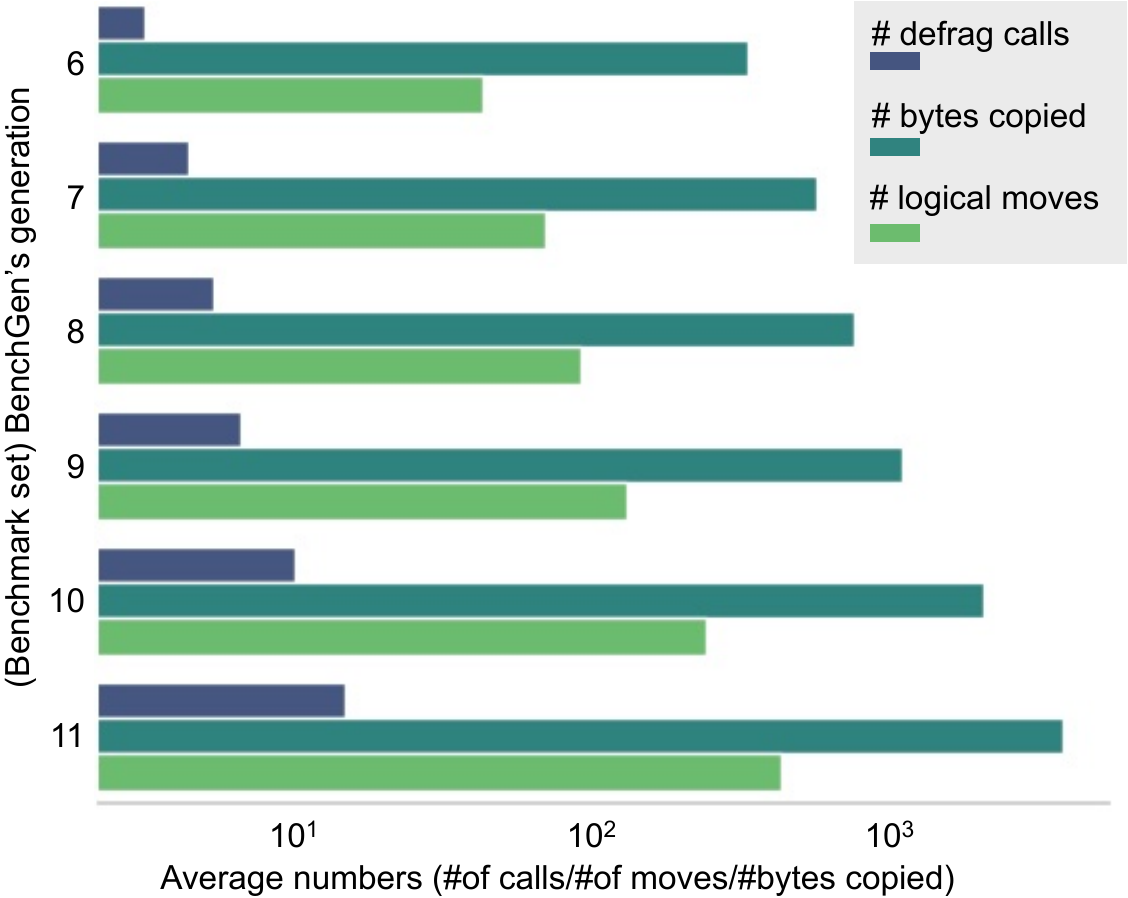}
\caption{Average number of calls to the defragmentation routine (Algorithm~\ref{alg_defragment}).}
\Description{Quantitative analysis of defragmentation (Algorithm~\ref{alg_defragment}).}
\label{fig_averageDefragmentationCalls}
\end{figure}

\paragraph{Discussion:}
Figure~\ref{fig_averageDefragmentationCalls} summarizes our observations.
First, defragmentation occurs infrequently.
In the 11th generation of benchmarks, we observed approximately 130 calls to Algorithm~\ref{alg_defragment} on average.
These benchmarks are large, containing more than 700{,}000 MLIR variables on average.

Each defragmentation call triggers, on average, between three and four memory moves, resulting in a total of 13 to 14 bytes moved per call.
This number is consistent with the median variable size, which lies between four and five bytes.
An average of four moves per call suggests that this is the typical number of overlapping live ranges at a program point.
Although the benchmarks are synthetic, this observation is consistent with prior studies on real programs.
For instance, \citet{Pereira05} report that most methods in the \texttt{java.util} library can be compiled using only four registers.

\subsection{RQ5: Running Time Impact}
\label{sub_rq5}

The main benefit of tree-scan allocation lies in reduced memory consumption, rather than running time.
However, loop unrolling and treeification impact the latter.
Unrolling has well-known performance benefits~\cite{Velkoski14}.
Additionally, we have observed that treeification, by eliminating join points, enables aggressive constant propagation and common subexpression elimination.
This section examines these effects.

\paragraph{Discussion:}
We did not observe statistically significant differences in execution time in the Honey Potion setting.
Using the Linux \texttt{bpftool} utility in profiling mode, we verified that all but two programs execute the same number of instructions, regardless of the allocator.
Note that loop unrolling is applied to every eBPF program by default, as it is required to generate code for the eBPF virtual machine.
Treeification, in turn, introduced minor changes in only two benchmarks, as discussed in Section~\ref{sub_rq1}.
Although tighter memory allocation might improve cache utilization, all benchmarks that successfully compile (two programs could not be compiled with the \textbf{trivial} allocator due to excessive stack requirement) use less than 512 bytes of stack space, which easily fits within the first-level cache of the target AMD processor.

\begin{figure}[ht]
\centering
\includegraphics[width=\columnwidth]{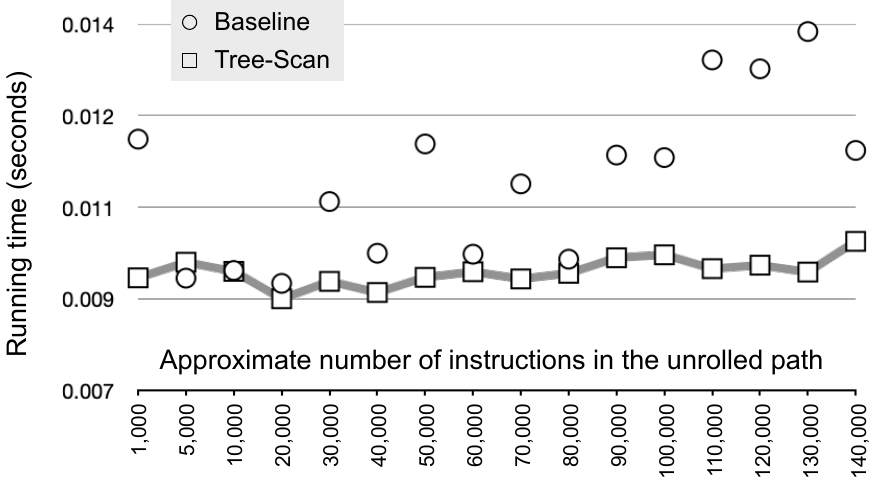}
\caption{Running time comparison between programs with and without treeification and tree-scan allocation, considering paths with different number of instructions.}
\Description{Running time comparison between programs with and without treeification and tree-scan allocation.}
\label{fig_runtimePath}
\end{figure}

To measure runtime differences, it is necessary to force the execution of long paths of treeified code.
Such experiments are only feasible in the MLIR setting.
Figure~\ref{fig_runtimePath} shows the execution time of paths of treeified code whose size varies from approximately 1,000 to 150,000 assembly instructions.
There is approximately one stack slot per three instructions.
The lifetimes of variables do not overlap; hence, tree-scan can keep all of them into a single stack slot.

\begin{figure}[ht]
\centering
\includegraphics[width=\columnwidth]{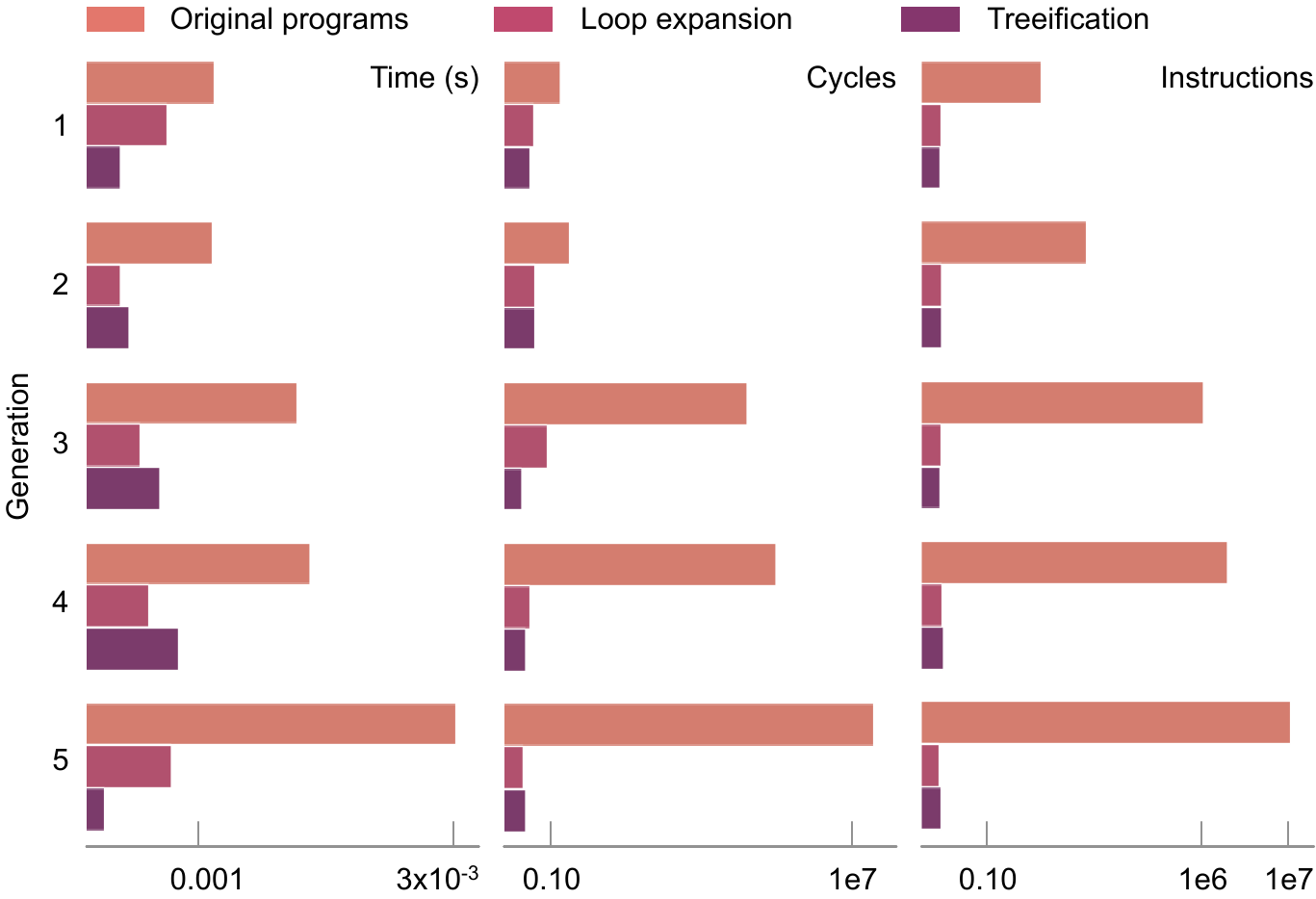}
\caption{Runtime of different generations of BenchGen programs. Time is measured with \texttt{hyperfine}. Number of cycles and instructions is measured with Linux \texttt{perf}.}
\Description{Runtime variation of different generations of BenchGen programs.}
\label{fig_performance_benchgen_programs}
\end{figure}

The treeified program incurs more cache misses: 1{,}557 vs.\ 719 in the instruction cache, and 14{,}827 vs.\ 10{,}838 in the data cache on the 150,000-instruction sample.
The additional data-cache misses are due to the loader touching a large amount of memory to relocate the expanded binary.
Nevertheless, the treeified programs run faster, and this difference increase as we increase the number of instructions in the executed path.
Figure~\ref{fig_performance_benchgen_programs} supports this statement: it shows the running time next to the number of cycles and number of instructions of five successive generations of BenchGen programs.
On the largest programs (fifth generation), we observe that the treeified program can run up to 30\% faster than the unrolled program, and up to 3.8x faster than the original program.

\subsection{RQ6: Impact of Treeification on Program Size}
\label{sub_rq6}

Treeification duplicates control-flow paths and, therefore, increases code size.
In the worst case, this growth is exponential in the number of branches in the program.
As observed in Section~\ref{sub_rq1}, treeification is rare in the eBPF setting, due to the simplicity of these programs.
To better expose its effects, we use BenchGen to generate programs with a high density of loops and branches.
This section analyzes the impact of treeification in this synthetic setting.

\begin{figure}[ht]
\centering
\includegraphics[width=\columnwidth]{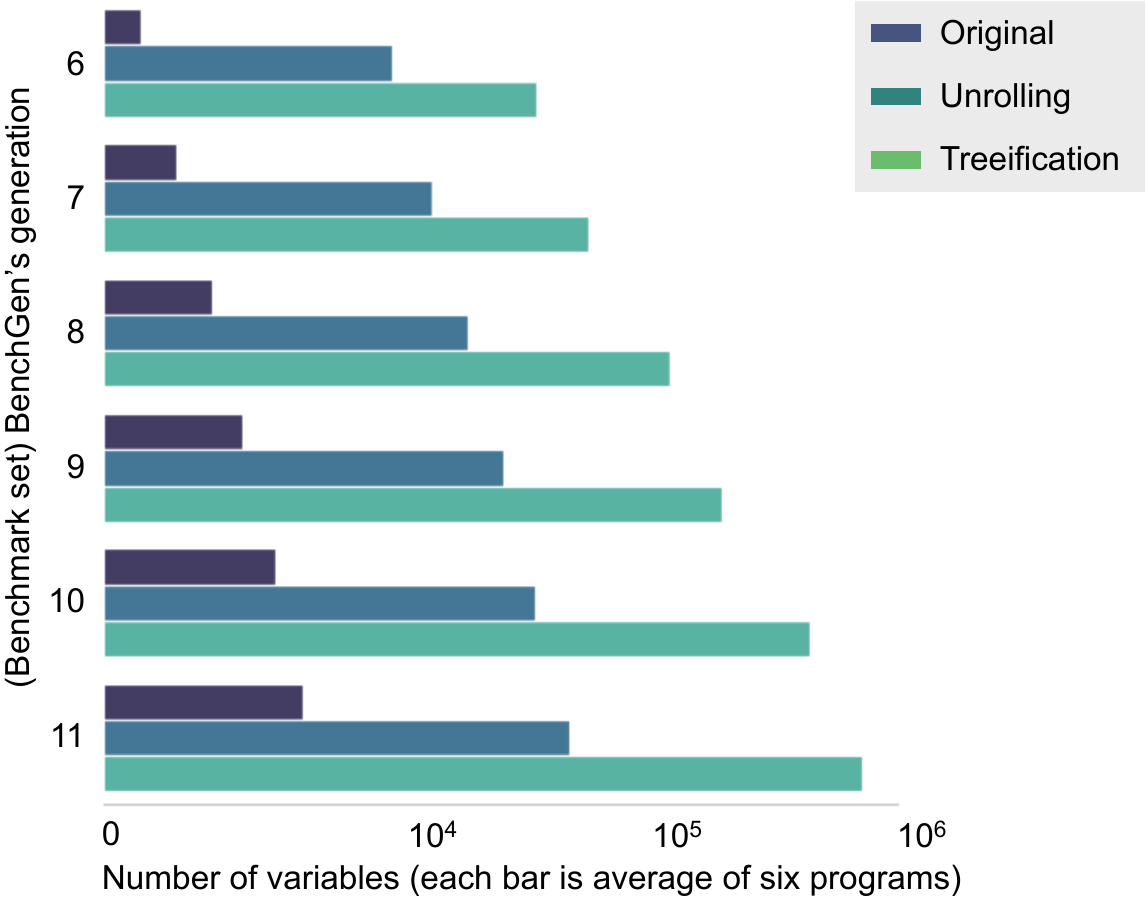}
\caption{Impact of treeification on code size.}
\Description{Impact of treeification on code size.}
\label{fig_averageProgramSize}
\end{figure}

\paragraph{Discussion:}
Figure~\ref{fig_averageProgramSize} shows the impact of treeification on code size.
The figure reports the number of variables at different stages of the pipeline shown in Figure~\ref{fig_runningCExample}, e.g.,
Fig.~\ref{fig_runningCExample} (A): original programs;
Fig.~\ref{fig_runningCExample} (B): programs after loop unrolling; and
Fig.~\ref{fig_runningCExample} (C): programs after treeification.
The number of instructions is proportional to the number of variables.
Code size grows rapidly: at the 11th generation, unrolling increases code size by almost 20x, and treeification increases the unrolled code by almost 10x.
Thus, in the MLIR setting, treeified programs are approximately 200x larger than the original programs.
However, we emphasize that these programs are synthetic.
Such growth was not observed in the real applications evaluated in Section~\ref{sub_rq1}.
Furthermore, as discussed in Section~\ref{sub_rq5}, treeified programs are not necessarily slower than their original versions; they are often faster due to the benefits of loop unrolling.

\section{Related Work}
\label{sec_rw}

This paper lies at the intersection of programming languages and algorithms, as it proposes a solution to a packing problem.
Variations of this problem have been studied before, and this section discusses this previous literature.

\paragraph{Dynamic Storage Allocation - No Defragmentation.}
The memory allocation problem that Section~\ref{sec_imp} solves has been studied in the past as the {\it Dynamic Storage Allocation} Problem~\cite{Robson74,Brent89,Knuth69} (DSA).
Given a set of tasks, each with a lifetime interval and a memory demand, DSA seeks to assign each task a contiguous range of memory addresses such that no two tasks that are simultaneously alive overlap in memory, while minimizing peak memory usage.
This problem corresponds to allocating weighted intervals on a line and admits a geometric interpretation as a two-dimensional packing problem.
DSA can be viewed as a restriction of the Storage Allocation Problem~\cite{BarYehuda13} (SAP), in which conflicts are defined over interval graphs rather than general interference graphs.
SAP itself is closely related to the \textit{Unsplittable Flow Problem on Paths} (UFPP)~\cite{BarYehuda13}, where tasks with demands must be assigned contiguous resources along a path subject to capacity constraints.
DSA is NP-complete~\cite{Garey02}, although constant-factor approximations are known, culminating in a 3-approximation for the general case~\cite{Gergov99}.
Previous work has proposed several heuristics for DSA, including greedy strategies such as First-Fit allocation, as well as techniques based on channel partitioning, block decomposition, and local rearrangements~\cite{Gergov96,Shuai04}.
None of these techniques consider the possibility of defragmenting memory.

\paragraph{Register Assignment - With Defragmentation.}
When defragmentation is allowed, the closest relatives to our packing problem arise in the context of register allocation.
General register allocation is NP-complete, as \citet{Chaitin81} showed via a reduction to graph coloring.
However, determining the minimum number of registers required to compile a program admits polynomial-time solutions when defragmentation is permitted.
In this setting, defragmentation corresponds to live-range splitting and register swapping.
\citet{Hack06} showed that register assignment can be solved in polynomial time for programs in static single assignment (SSA) form.
Programs in SSA form have chordal interference graphs~\cite{Pereira05}, which can be colored in polynomial time.
In the absence of swaps, minimizing the number of registers remains NP-complete~\cite{Pereira06,Bouchez06}.
When registers have different sizes (e.g., 32-bit and 64-bit), and the compiler can pack smaller variables into larger registers~\cite[Sec.2.1]{VenkataKeerthy23}, minimizing the number of registers remains NP-complete~\cite{Lee07,Lee08}.
Interestingly, if register swapping is allowed, the problem becomes polynomial-time again for architectures with single-, double-, and quadruple-precision registers~\cite{Pereira08}.
These formulations assume a finite set of register sizes.
In hardware synthesis, this restriction can be relaxed: minimizing the total number of bits required to implement a register bank can be solved in polynomial time, as shown by \citet{Canesche22}, again assuming that register swapping is available.
The packing problem studied in this paper, i.e., Dynamic Storage Allocation with Defragmentation, differs from these formulations.
As an intuitive analogy, one may assume an unbounded bank of registers, where each register can be subdivided and contiguous regions can be combined.

\section{Conclusion}
\label{sec_conc}

This paper has introduced a static memory allocator for constant-bounded programs, that is, programs whose number of operations can be determined from their syntax.
The allocator achieves a minimal memory footprint by duplicating program paths, greedily assigning memory regions to variables, and defragmenting memory when necessary.
Memory assignment is fully static, even in the context of stack or heap allocation.

Although the approach may increase code size, potentially leading to exponential growth, the paper identifies a practical use case: the compilation of eBPF programs.
We observe that these programs are typically simple enough that neither code duplication nor defragmentation is likely to occur in practice.
Furthermore, we evaluate the behavior of the proposed allocator on more complex programs, showing that for moderately sized inputs, static tree-scan allocation with code duplication remains practical.


\section*{Acknowledgment}
This project was supported by FAPEMIG (Grant APQ-00440-23), CNPq (Grant \#444127/2024-0), CAPES (\textsc{PrInt}), and Google, which sponsored Vin\'{i}cius Silva's scholarship.  We are grateful to Xinliang (David) Li and Victor Lee for their efforts in making the Google sponsorship possible.

\bibliographystyle{ACM-Reference-Format}
\bibliography{references}

\end{document}